\documentclass[twocolumn,showpacs,preprintnumbers,prd,superscriptaddress,nofootinbib]{revtex4-1}

\usepackage{amsmath}
\usepackage{amsfonts}
\usepackage{amssymb}
\usepackage{graphicx}
\usepackage{color}
\usepackage{hyperref}
\usepackage{booktabs}
\usepackage{hyperref}
\usepackage{cleveref}
\usepackage{color}

\Crefname{equation}{Eq.}{Eqs.}
\Crefname{figure}{Fig.}{Figs.}
\Crefname{section}{Sec.}{Secs.}

\usepackage{etoolbox}
\makeatletter
\appto{\appendix}{%
  \@ifstar{\def\theequation@prefix{A.}}%
          {}%
}
\makeatother

\begin{document}

\title{Cosmic acceleration from a single fluid description}

\author{Salvatore Capozziello}
\email{capozzie@na.infn.it}
\affiliation{Dipartimento di Fisica, Universit\`a di Napoli  ``Federico II'', Via Cinthia, I-80126, Napoli, Italy.}
\affiliation{Istituto Nazionale di Fisica Nucleare (INFN), Sez. di Napoli, Via Cinthia 9, I-80126 Napoli, Italy.}
\affiliation{Gran Sasso Science Institute, Via F. Crispi 7, I-67100, L' Aquila, Italy.}

\author{Rocco D'Agostino}
\email{rocco.dagostino@roma2.infn.it}
\affiliation{Dipartimento di Fisica, Universit\`a degli Studi di Roma ``Tor Vergata'', Via della Ricerca Scientifica 1, I-00133, Roma, Italy.}
\affiliation{Istituto Nazionale di Fisica Nucleare (INFN), Sez. di Roma ``Tor Vergata'', Via della Ricerca Scientifica 1, I-00133, Roma, Italy.}

\author{Orlando Luongo}	
\email{orlando.luongo@lnf.infn.it}
\affiliation{Istituto Nazionale di Fisica Nucleare, Laboratori Nazionali di Frascati, 00044 Frascati, Italy.}
\affiliation{School of Science and Technology, University of Camerino, I-62032, Camerino, Italy.}
\affiliation{Department of Mathematics and Applied Mathematics, University of Cape Town, Rondebosch 7701,
Cape Town, South Africa.}
\affiliation{Astrophysics, Cosmology and Gravity Centre (ACGC), University of Cape Town, Rondebosch 7701,
Cape Town, South Africa.}


\begin{abstract}
We here propose a new class of barotropic factor for matter, motivated by properties of isotropic deformations of crystalline solids. Our approach is dubbed Anton-Schmidt's equation of state and provides a non-vanishing, albeit small, pressure term for matter. The corresponding pressure is thus proportional to the logarithm of universe's volume, i.e. to the density itself since $V\propto \rho^{-1}$. In the context of solid state physics, we demonstrate that by only invoking standard matter with such a property, we are able to frame the universe speed up in a suitable way, without invoking a dark energy term by hand. Our model extends a recent class of dark energy paradigms named \emph{logotropic} dark fluids and depends upon two free parameters, namely $n$ and $B$. Within the Debye approximation, we find that $n$ and $B$ are related to the Gr\"uneisen parameter and the bulk modulus of crystals.
We thus show the main differences between our model and the logotropic scenario, and we highlight the most relevant properties of our new equation of state on the background cosmology. Discussions on both kinematics and dynamics of our new model have been presented. We demonstrate that the $\Lambda$CDM model is inside our approach, as limiting case. Comparisons with CPL parametrization have been also reported in the text.
Finally, a Monte Carlo analysis on the most recent low-redshift cosmological data allowed us to place constraints on $n$ and $B$. In particular, we found $n=-0.147^{+0.113}_{-0.107}$ and $B=3.54 \times 10^{-3}$.
\end{abstract}

\maketitle


\section{Introduction}

In the context of homogeneous and isotropic universe, Einstein's gravity supplies a cosmological description which commonly makes use of a perfect fluid energy momentum tensor, with barotropic equation of state given by the ratio between pressure and density \cite{uno1,uno2}. Observations have shown that under this hypothesis a negative anti-gravitational equation of state is requested to speed up the universe at our times \cite{supernovae,Suzuki11}. Several scenarios have been introduced in the literature with the aim of describing the cosmic acceleration in terms of exotic fluids which counterbalance the action of gravity \cite{due1,due2}. The simplest approach is offered by a quantum vacuum energy density given by the cosmological constant $\Lambda$ \cite{costante}. One of the advantages of $\Lambda$ is to provide a constant equation of state, i.e. $w=-1$, with constant pressure and density. Although attractive, the corresponding paradigm, named the $\Lambda$CDM model, is jeopardized by some caveats which may limit its use. Dark energy is the simplest class of extensions of the concordance model. Even though a wide number of dark energy models has been introduced \cite{Copeland06,Bamba12}, the problem of the onset and nature of cosmic acceleration remains an open challenge of modern cosmology \cite{altro1}.

\noindent Among all approaches, modifying the fluid responsible for the cosmic acceleration is likely the simplest way to account for the dark energy properties at large scales. A relevant example has been offered by Chaplygin gas \cite{chap1} and extensions which are built up with different equations of state \cite{chap2}. Since the cosmic pressure is negative, we wonder whether matter alone can be used to provide, at certain stages of universe's evolution, a negative pressure \cite{altro2}. To do so, we analyze in nature possible cases in which this may happen. One possibility is to consider standard matter with a different equation of state compared to the usual case in which its pressure vanishes. The process that enables matter to pass from a pressureless equation of state to a negative pressure is due to the cosmic expansion as the standard model provides.

We thus follow this strategy and we propose matter obeying the Anton-Schmidt's equation of state within the Debye approximation \cite{AntonSchmidt}. The aforementioned framework has been introduced to empirically describe the pressure of crystalline solids which deform under isotropic stress (for a review, see \cite{bis}). If one considers the universe to deform under the action of cosmic expansion, it would be possible to model the fluid by means of such an equation of state that turns out to be naturally negative \cite{bis2}.

In this paper, we apply the properties of Anton-Schmidt's equation of state in the framework of Friedmann-Robertson-Walker (FRW) universe. We first motivate the choice of Anton-Schmidt's equation of state. Afterwards, we show the technique to pass from pressureless matter to Anton-Schmidt dark energy scenario. Here, we show that if matter obeyed the Anton-Schmidt's equation of state, it would be possible to fuel the universe to accelerate without the need of the cosmological constant, i.e. $\Lambda$. To figure this out, we highlight that dark energy becomes a consequence of Anton-Schmidt's equation of state in which matter naturally provides a negative barotropic factor. This unusual behaviour for matter is the basic demand to get speed up after the transition redshift. We analyze different epochs, i.e. at late-times and before current era, and we identify the mechanism responsible for the cosmic acceleration. We investigate some properties of our model and we show that it depends upon two parameters only, characterized by precise physical meanings. The key feature lies on analyzing the role played by universe's volume which influences the equation of state itself. In particular, a negative value for pressure becomes dominant as the volume of the universe overcomes a certain value. We thus show that although the model drives the universe dynamics at all stages, relevant consequences raise only as the volume takes a given value, enabling the pressure not to be negligible. In this scenario, we notice our model matches the approach of logotropic dark energy (LDE) model \cite{Chavanis15}. We thus propose a common origin between Anton-Schmidt dark energy and LDE. To do so, we investigate the limits to LDE and we show in which regimes our scenario becomes equivalent, emphasizing the main differences between the two approaches. To consolidate our theoretical alternative to $\Lambda$, we finally compute observational constraints at the level of background cosmology using supernova data, differential Hubble rate and baryon acoustic oscillation measurements. We thus perform a combined fit providing numerical bounds over our free parameters by means of Monte Carlo analysis.

The paper is structured as follows. In \Cref{sec:due}, we describe the Anton-Schmidt scenario for crystalline solids in the Debye approximation. We also emphasize the role played by $V_0$, as scaling volume inside the pressure definition. Afterwards, in the same section we discuss the technique able to split the whole energy density into two parts, corresponding to a pressureless term and a pushing up effective dark energy counterpart. In \Cref{sec:tre}, we thus describe how acceleration born from our framework, highlighting implications in modern cosmology and then comparing our outcomes with the concordance and the Chevallier-Polarski-Linder parametrization. We then report the onset of cosmic acceleration and the role played by the sound speed. In \Cref{sec:constraints}, we describe observational constraints over our model, which lie on the Monte Carlo techniques based on affine-invariant ensemble sampler. We conclude our work with conclusions and perspectives summarized in \Cref{sec:conclusions}.


\section{Anton-Schmidt's matter fluid as effective dark energy}
\label{sec:due}

The issue of negative pressure jeopardizes the cosmological standard model, since it is not easy to get evidences of negative pressure in laboratory. However, in the framework of condensed matter and solid state physics it could happen that effective pressures become locally negative. Some other cases permit scenarios in which the physical counterpart of the pressure is effectively negative. This is the case of the Anton-Schmidt's equation of state for crystalline solids \cite{AntonSchmidt}. In particular, the Anton-Schmidt's equation of state gives the empirical expression of crystalline solid's pressure \emph{under isotropic deformation}. In the Debye approximation, one can write the Anton-Schmidt's pressure as follows:
\begin{equation}
P(V)=-\beta\left(\dfrac{V}{V_0}\right)^{-\frac{1}{6}-\gamma_G}\ln\left(\dfrac{V}{V_0}\right)\,,
\label{eq:Anton-Schmidt}
\end{equation}
where

\begin{itemize}
\item $V$ is the volume of the crystal, and  $V_0$ is the equilibrium volume. In particular, $V_0$ shows the limit at which the pressure vanishes. This occurrence is found as $V=V_0$, enabling $V_0$ to be considered as a \emph{barrier} which shifts among different signs of $P$. In fact, as $V<V_0$ the pressure is positive, for positive bulk modulus, while negative in the opposite case, i.e. $V_0<V$. For negative bulk modulus, the cases are reversed, although negative $\beta$ are here excluded since apparently not significant in our analyses;
\item $\beta$ is the bulk modulus at $V_0$. This quantity is intimately related to sound perturbations and elasticity in fluids. In such a picture, $\beta$ is in analogy with the spring constant of an oscillator, once given a fluid or a crystal. It may be viewed as a heuristic measurement of how much physical dimensions, i.e. volume, lengths, change under the action of external forces. In our case, we take the standard definition of $\beta$ and we assume it is related to the variation of $P$ in terms of the volume by
 \begin{equation}
\beta=-V_0\left(\dfrac{dP}{dV}\right)_{V=V_0}\ ;
 \end{equation}
\item $\gamma_G$ is the dimensionless Gr\"uneisen parameter \cite{Gruneisen12}, which has both a thermal and an equivalent microscopic interpretation. The macroscopic definition is related to the thermodynamic properties of the material:
\begin{equation}
\gamma_G=\frac{\alpha V K_T}{C_V}\ ,
\end{equation}
where $\alpha$ is the thermal coefficient, $K_T$ is the isothermal bulk modulus\footnote{It is a widely-accepted convention to refer to the bulk modulus as $\beta$ and to its isothermal version as $K_T$, instead of a likely more immediate $\beta_T$.}, and $C_V$ is the heat capacity at constant volume. The microscopic formulation accounts for the variation of the vibrational frequencies of the atoms in the solid with $V$. In fact, the Gr\"uneisen parameter of an individual vibrational mode $i$ is given by
\begin{equation}
\gamma_i=-\dfrac{d \ln \omega_i}{d \ln V}\ ,
\end{equation}
where $\omega_i$ is the vibrational frequency of the $i$th mode.
Under the quasi-harmonic approximation, it is possible to relate the macroscopic definition of $\gamma_G$ to its microscopic definition  if one writes \cite{Barron57}
\begin{equation}
\gamma_G=\dfrac{1}{C_V}\sum_i C_{V,i} \gamma_i = \dfrac{\alpha V K_T}{C_V}\ ,
\end{equation}
where $C_{V,i}$ is the contribution of each mode to the heat capacity. In the Debye model, the Gr\"uneisen parameter reads
\begin{equation}
\gamma_{G}=\dfrac{d \ln \theta_D}{d \ln V}	\ ,
\end{equation}
where $\theta_D$ is the Debye temperature \cite{Debye} defined as $\theta_D=	\hbar \omega_D/k_B$, where $\hbar$ and $k_B$ are the Planck's and Boltzmann's constants, respectively, and $\omega_D$ is the maximum vibrational frequency of a solid's atoms.
The values of the Gr\"uneisen parameter do not show big variations for a wide variety of chrystals \cite{Anderson89} and are typically in the range 1 to 2.
\end{itemize}

As a consequence of our recipe, under the hypothesis of Anton-Schmidt fluid, we immediately find that:

\begin{enumerate}
  \item the whole universe may be modelled by a single dark counterpart under the form of Anton-Schmidt fluid. In particular, one can consider that matter fuels the cosmic speed up, if its equation of state is supposed to be the one of \Cref{eq:Anton-Schmidt};
  \item admitting that matter depends on Anton-Schmidt's equation of state means that $P\neq0$ for matter. So the form of matter distribution in the whole universe is not exactly zero;
  \item assuming for matter that $P\neq0$, under the form of \Cref{eq:Anton-Schmidt}, is equivalent to employ a non-vanishing equation of state for matter, but small enough to accelerate the universe with a negative sign;
  \item the sign of Anton-Schmidt's equation of state for matter is naturally negative, by construction of the pressure itself;
  \item the parameter $\gamma_G$ is not arbitrary and depends on the kind of fluid entering the energy momentum tensor. In the case of matter, in the homogeneous and isotropic universe, $\gamma_G$ will be a free parameter of the theory itself.
\end{enumerate}

In our case, since we only have matter obeying \Cref{eq:Anton-Schmidt}, to guarantee the cosmic speed up at late times, one needs to overcome the limit:
\begin{equation}\label{barriera}
\gamma_G=-\frac{1}{6}\,.
\end{equation}
If $\gamma_G>-\frac{1}{6}$ the cosmic acceleration does not occur at $z\simeq0$. In all cases the advantages of employing a matter term evolving as \Cref{eq:Anton-Schmidt} are:
\begin{itemize}
  \item the pressure \emph{is not postulated to be negative a priori} as in the standard cosmological model. Notice that in the $\Lambda$CDM case one has at least two fluids: the first concerning pressureless matter, whereas the second composed by $\Lambda$. In such a picture, what pushes up the universe to accelerate is the cosmological constant. In our puzzle, neglecting all small contributions, such as neutrinos, radiations, spatial curvature, etc., one finds that standard matter is enough to enable the acceleration once Eq. \eqref{eq:Anton-Schmidt} is accounted;
  \item the physical mechanism behind Eq. \eqref{eq:Anton-Schmidt} states that if the universe expands then the net equation of state provides different behaviours, corresponding to deceleration at certain times and  acceleration during other epochs;
  \item it is possible to measure in a laboratory the effects of \Cref{eq:Anton-Schmidt}, which is physical and does not represent a \emph{ad hoc} construction of the universe pressure.
\end{itemize}

It is natural to suppose that \Cref{eq:Anton-Schmidt} bids to the following limits:

\begin{eqnarray}
\left\{
  \begin{array}{ll}
    \lim_{V\gg V_0}\,\,\Big|P(V)\Big|=\infty, & \hbox{$\gamma_G<-\dfrac{1}{6}$;} \nonumber\\
    \,\\
    \lim_{V\ll V_0}\,\,\Big|P(V)\Big|=0\,, & \hbox{$\gamma_G<-\dfrac{1}{6}$.}\nonumber
  \end{array}
\right.
\end{eqnarray}

Thus, there exists a volume at which the matter pressure  turns out to be dominant over the case $P=0$. It follows that matter with the above pressure can accelerate the universe after a precise time. This is a consequence of our model and it is not put by hand as in the concordance paradigm.

Indeed, introducing \emph{two fluids}: matter and $\Lambda$ there exists a time at which $\Lambda$ dominates over matter and pushes up the universe. In our approach, there exists \emph{one fluid}, with a single equation of state, able to accelerate the universe as the volume passes the barrier $V\simeq V_0$. This means that the transition redshift is not actually relevant, because having one fluid only, the whole universe dynamics is essentially dominated by the fluid dynamics itself.

We can focus on three different cases, reported below.
\begin{description}
  \item[case 1] The time before passing the $V_0$ barrier, i.e. $V<V_0$. In such a case, there exists an expected matter dominated phase. Indeed, when $V<V_0$, one gets the limit above stated which provides exactly the pressureless case $P=0$, as in the standard model paradigm.
  \item[case 2] the time of equivalence between $V$ and the $V_0$ barrier, i.e. $V=V_0$. In this case, we lie on the transition time, which occurs as $V=V_0$. In this case there exists a transition at a transition time, which leads to $V(z_{tr})=V_0$.
  \item[case 3] the time after passing the $V_0$ barrier. Here, since $V>V_0$ one passes from $P=0$ to $P<0$ and matter starts to accelerate the universe instead of decelerating it.
\end{description}

Following the notation introduced in the context of the LDE model \cite{Chavanis15}, we express the volume in terms of mass density, $V\propto \rho^{-1}$, and recast \Cref{eq:Anton-Schmidt} as
\begin{equation}
P(\rho)=A\left(\dfrac{\rho}{\rho_*}\right)^{-n}\ln\left(\dfrac{\rho}{\rho_*}\right)\ ,
\label{eq:Anton-Schmidt 2}
\end{equation}
where $\rho_*$ is a reference density\footnote{$\rho_*$ has been identified with the Planck density in \cite{Chavanis15}: $\rho_P=c^5/\hbar G^2\approx 5.16\times 10^{99}$ g/m$^3$. The physical motivation and the implications of this choice will be discussed in \Cref{sec:constraints}.}. The new notation implies $A\propto \beta$ and $n=-\frac{1}{6}-\gamma_G$.
For $n=0$, \Cref{eq:Anton-Schmidt 2} reduces to the equation of state characteristic of the logotropic cosmological models \cite{Chavanis15}, in which the constant $A$ represents the logotropic temperature, positive-definite.

In the present work we want to study the dynamical evolution of a universe made of a single fluid described by the Anton-Schmidt's equation of state. We assume homogeneity and isotropy on large scales \cite{Peebles93} and, in agreement with the Cosmic Microwave Background (CMB) observations \cite{Planck15}, we consider a flat universe. Then, the Friedmann equation is given by
\begin{equation}
\left(\dfrac{\dot a}{a}\right)^2=\dfrac{8\pi G}{3c^2}\epsilon\ ,
\label{eq:Friedmann}
\end{equation}
where the `dot'  indicates derivative with respect to the cosmic time $t$, and $a(t)$ is the scale factor normalized to unity today  $(a_0=1)$
To determine the total energy density $\epsilon$, we assume an adiabatic evolution for the fluid, so that the first law of thermodynamics reads
\begin{equation}
d\epsilon=\left(\dfrac{\epsilon+P}{\rho}\right)d\rho\ .
\label{eq:first law}
\end{equation}
For $P(\rho)$ as given in \Cref{eq:Anton-Schmidt 2}, the above equation can be integrated and one obtains
\begin{align}
\epsilon&=\rho c^2+\rho\int^{\rho} d\rho' \ \dfrac{p(\rho')}{\rho'^2} \label{eq:total energy} \\
&=\rho c^2-\left[\dfrac{A}{n+1}\left(\dfrac{\rho}{\rho_*}\right)^{-n}\ln\left(\dfrac{\rho}{\rho_*}\right)+\dfrac{A}{(n+1)^2}\left(\dfrac{\rho}{\rho_*}\right)^{-n}\right],  \nonumber
\end{align}
for $n\neq-1$. \Cref{eq:total energy} tells us that the energy density of the fluid is the sum of its rest-mass energy ($\rho c^2$) and its internal energy. The first term, describing a pressureless fluid, mimics matter, while the second term, which arises from pressure effects, can be interpreted as the dark energy term. Hence, this approach intends to unify dark matter and dark energy into a single dark fluid. Nonetheless, to draw a parallel with the standard scenario in which matter and dark energy are expressions of two separate fluids, we decide to write \Cref{eq:total energy} as
\begin{equation}
\epsilon=\epsilon_m+\epsilon_{de}\,,
\label{eq:epsilon}
\end{equation}
with
\begin{align}
&\epsilon_m=\rho c^2\ , \label{eq:e_m}\\
&\epsilon_{de}=-\dfrac{A}{n+1}\left(\dfrac{\rho}{\rho_*}\right)^{-n}\ln\left(\dfrac{\rho}{\rho_*}\right)-\dfrac{A}{(n+1)^2}\left(\dfrac{\rho}{\rho_*}\right)^{-n}. \label{eq:e_de}
\end{align}

In the early universe ($a\rightarrow 0$, $\rho \rightarrow \infty$), the rest-mass energy $\epsilon_m$ dominates. However, if $n<0$  the pressure given in \Cref{eq:Anton-Schmidt 2} is not vanishing as expected in the standard matter-dominated universe. This is due to the fact that the Anton-Schmidt's equation of state in the Debye approximation cannot describe the cosmic fluid of the early phase cosmology, when temperatures are much higher than the Debye temperatures of solids. In any case, if one wants to extend our model to early times would notice that $P\ll \epsilon$ for $\rho \rightarrow \infty$, and thus the fluid behaves as if it were pressureless,  similarly to the case of the LDE model, as pointed out in \cite{Chavanis17}. At late times $(\rho\ll 1)$, the internal energy $\epsilon_{de}$ dominates and, for $n<0$,  the pressure tends to a constant negative value as in a universe that is dominated by dark energy.

In the next section, we shall analyze the cosmological implications of such a picture at the background level. We compare the features of the model we propose with the standard cosmological scenarios, and we investigate the epoch of the dark energy dominance that drives the accelerated expansion of the universe.

\section{Consequences on background cosmology}
\label{sec:tre}

\begin{figure}[h!]
\begin{center}
\includegraphics[width=3.in]{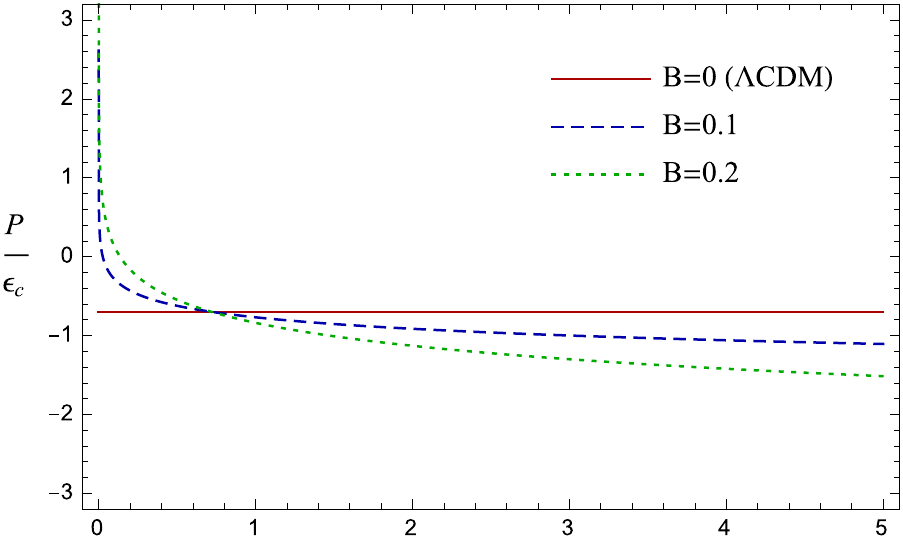}\\
\vspace{0.2cm}
\includegraphics[width=3.in]{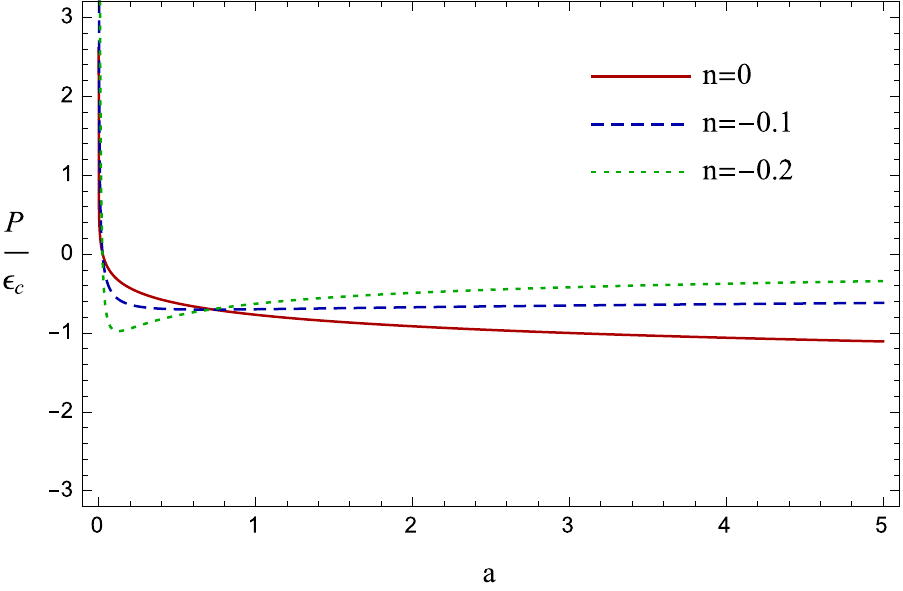}
\caption{Dynamical evolution of the pressure for different values of $B$ and $n=0$ (\textit{top}), and for different values of $n$ and $B=0.1$ (\textit{bottom}). The matter density parameter is fixed to $\Omega_{m0}=0.3$.}
\label{fig:p(a)}
\end{center}
\end{figure}

\begin{figure}[h!]
\begin{center}
\includegraphics[width=3.in]{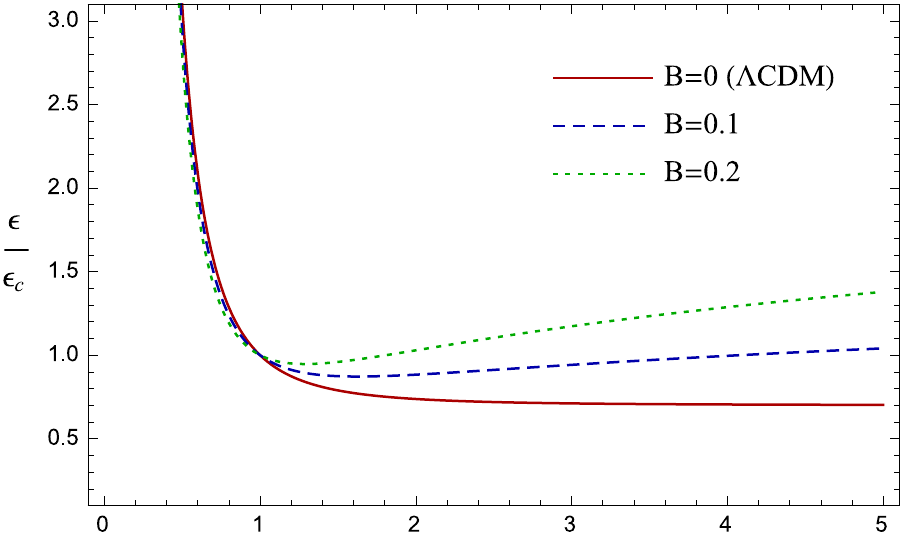}\\
\vspace{0.2cm}
\includegraphics[width=3.in]{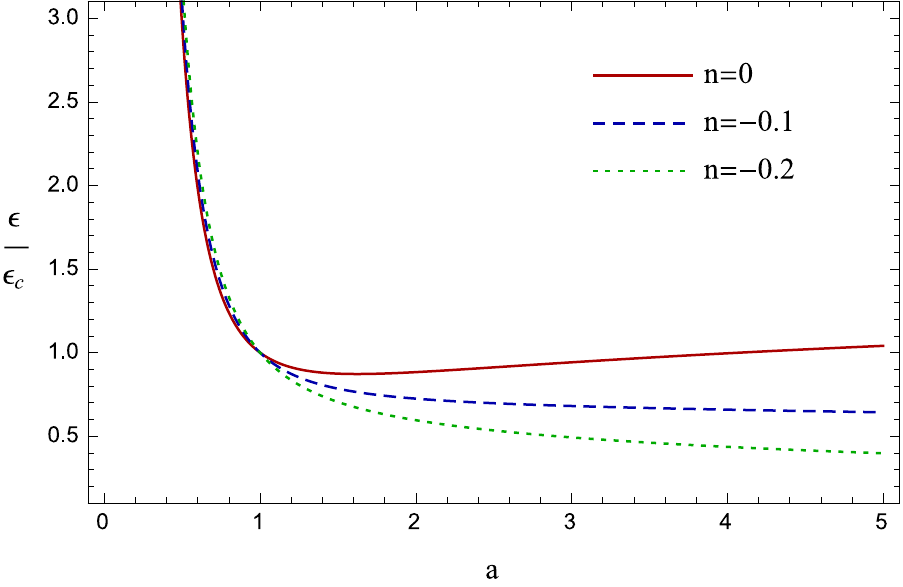}
\caption{Dynamical evolution of the energy density for different values of $B$ and $n=0$ (\textit{top}), and for different values of $n$ and $B=0.1$ (\textit{bottom}). $\Omega_{m0}$ is kept fixed to the value of 0.3.}
\label{fig:e(a)}
\end{center}
\end{figure}

\begin{figure}[h!]
\begin{center}
\includegraphics[width=3.in]{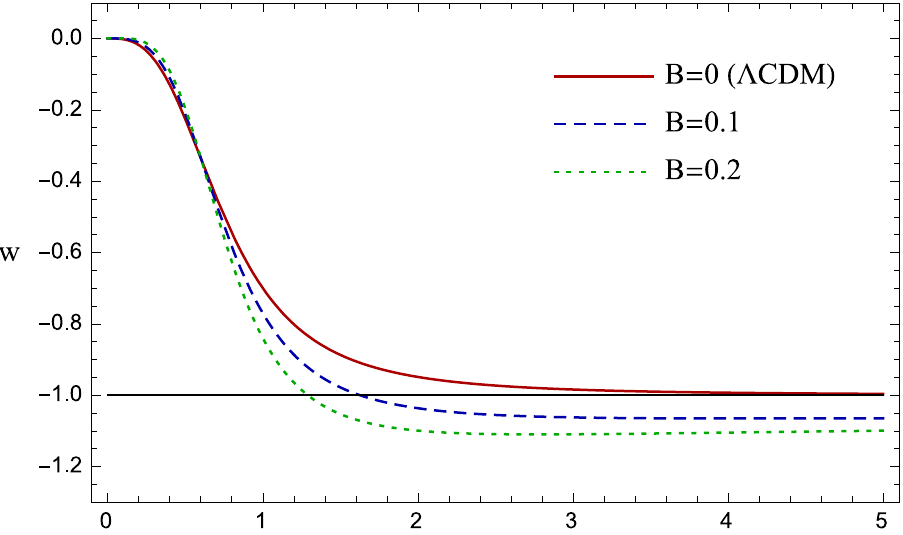}\\
\vspace{0.2cm}
\includegraphics[width=3.in]{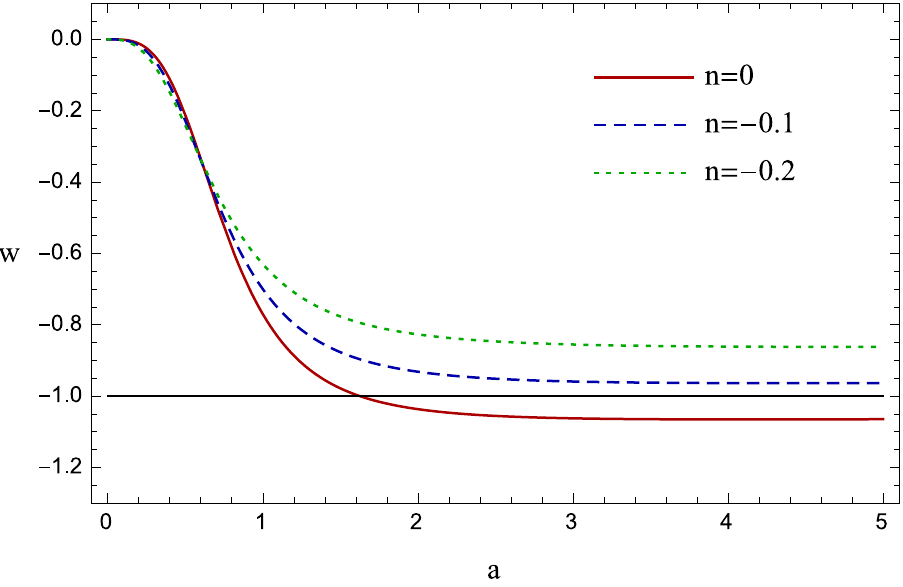}
\caption{Dynamical evolution of the equation of state parameter for different values of $B$ and $n=0$ (top), and for different values of $B$ and $n=0$. In both plots, $\Omega_{m0}=0.3$. The black line separates the ``normal" values from the phantom values $(w<-1)$.}
\label{fig:w(a)}
\end{center}
\end{figure}

The relation between the energy density and the scale factor for a given barotropic fluid is given by the continuity equation:
\begin{equation}
\dot{\epsilon}+3\dfrac{\dot{a}}{a}(\epsilon+P)=0\ .
\label{eq:continuity}
\end{equation}
Combining \Cref{eq:first law} and \Cref{eq:continuity} yields
\begin{equation}
\dot{\rho}+3\dfrac{\dot{a}}{a}\rho=0\,,
\end{equation}
which, once integrated, gives the evolution of the rest-mass density:
\begin{equation}
\rho=\dfrac{\rho_{m,0}}{a^{3}}\ .
\label{eq:matter}
\end{equation}
Using \Cref{eq:matter,eq:e_m,eq:e_de}, we get
\begin{align}
&\epsilon_m=\dfrac{\rho_{m,0} c^2}{a^3}, \label{eq:e_m1}\\
&\epsilon_{de}=-\dfrac{A}{n+1}\left(\dfrac{\rho_{m,0}}{\rho_*a^3}\right)^{-n}\ln\left(\dfrac{\rho_{m,0}}{\rho_*a^3}\right)-\dfrac{A}{(n+1)^2}\left(\dfrac{\rho_{m,0}}{\rho_*a^3}\right)^{-n}, \label{eq:e_de1}
\end{align}
or equivalently,
\begin{align}
&\epsilon_m=\epsilon_{m,0}a^{-3}, \label{eq:e_m2}\\
&\epsilon_{de}=\epsilon_{de,0}a^{3n}+\dfrac{3A}{n+1}\left(\dfrac{\rho_{m,0}}{\rho_*}\right)^{-n}a^{3n}\ln a, \label{eq:e_de2}
\end{align}
where $\epsilon_{m,0}$ and $\epsilon_{de,0}$ are the energy densities evaluated at the present time.
Introducing the Hubble parameter $H\equiv\dot{a}/a$ and the critical energy density $\epsilon_c\equiv3H_0^2c^2/(8\pi G)$, \Cref{eq:Friedmann} can be rewritten as\footnote{The contribution of radiation is neglected at late times $(a\sim 1)$.}
\begin{equation}
H^2=H_0^2\left(\dfrac{\epsilon_m}{\epsilon_c}+\dfrac{\epsilon_{de}}{\epsilon_c}\right) ,
\label{eq:Hubble}
\end{equation}
where the energy density has been decomposed as in \Cref{eq:epsilon}. One can define the normalized matter and dark energy densities,
\begin{subequations}
\begin{align}
\Omega_{m0}&\equiv \dfrac{\epsilon_{m,0}}{\epsilon_c}\,, \label{eq:Omega_m}\\
\Omega_{de,0}&\equiv \dfrac{\epsilon_{de,0}}{\epsilon_c}\,,
\end{align}
\end{subequations}
satisfying the condition $\Omega_{de,0}=1-\Omega_{m0}$. Thus, after simple manipulations, \Cref{eq:Hubble} becomes
\begin{equation}
H^2=H_0^2\left[\dfrac{\Omega_{m0}}{a^3}+(1-\Omega_{m0})(1+3B\ln a)a^{3n}\right]\,,
\label{eq:Hubble1}
\end{equation}
where, for convenience, we introduce
\begin{equation}
B=\dfrac{A}{n+1}\left(\dfrac{\rho_{m,0}}{\rho_*}\right)^{-n}\dfrac{1}{\epsilon_c\Omega_{de,0}}\ .
\label{eq:B parameter}
\end{equation}
For $n=0$, \Cref{eq:B parameter} reduces to
\begin{equation}
B\big|_{n=0}=\dfrac{A}{\epsilon_c\Omega_{de,0}}\ ,
\end{equation}
which represents the dimensionless logotropic temperature. To calculate the equation of state parameter $w=P/\epsilon$
in terms of the logotropic temperature, we rearrange \Cref{eq:Anton-Schmidt 2,eq:epsilon} as
\begin{align}
&P=-\epsilon_c(1-\Omega_{m0})\left[B+(n+1)(1+3B\ln a)\right]a^{3n} \label{eq:p} , \\
&\epsilon=\dfrac{\epsilon_c\Omega_{m0}}{a^3}+\epsilon_c(1-\Omega_{m0})(1+3B\ln a)a^{3n}  \label{eq:e}\ .
\end{align}
Thus, one obtains
\begin{equation}
w=-\dfrac{(1-\Omega_{m0})\left[B+(n+1)(1+3B\ln a)\right]a^{3n}}{\Omega_{m0}a^{-3}+(1-\Omega_{m0})\left(1+3B\ln a\right)a^{3n}}\ .
\label{eq:w}
\end{equation}
Fixing the indicative value $\Omega_{m0}=0.3$, we show in \Cref{fig:p(a)} the behaviour of $P$  for different values of the parameters $B$ and $n$. \Cref{fig:e(a)} shows the evolution of the energy density in terms of the scale factor. For $B>0$, after reaching the minimum, the energy density increases with the scale factor characterizing a phantom universe. The behaviour of the equation of state parameter is shown in \Cref{fig:w(a)}.

For $a\rightarrow \infty$, $w\rightarrow -1$.
When $n=0$, the equation of state parameter reduces to
\begin{equation}
w\big|_{n=0}=-\dfrac{(1-\Omega_{m0})(B+1+3B\ln a)}{\Omega_{m0}a^{-3}+(1-\Omega_{m0})(1+3B\ln a)}\ ,
\end{equation}
proper of the LDE model \cite{Chavanis15}.

We now focus our study on the dark energy equation of state parameter: $w_{de}=P_{de}/\epsilon_{de}$.
Since the contribution of matter is negligible, one can simply identify the total pressure given in \Cref{eq:p} with the dark energy pressure,
\begin{equation}
P_{de}=-\epsilon_c(1-\Omega_{m0})\left[B+(n+1)(1+3B\ln a)\right]a^{3n},
\label{eq:p_de}
\end{equation}
while the dark energy density term reads (cf. \Cref{eq:e})
\begin{equation}
\epsilon_{de}=\epsilon_c(1-\Omega_{m0})(1+3B\ln a)a^{3n}\ .
\end{equation}
Thus, one finally obtains
\begin{equation}
w_{de}=-(n+1)-\dfrac{B}{1+3B\ln a}\ .
\label{eq:dark energy EoS}
\end{equation}


\subsection{Comparison with the concordance model}

Our results contain the concordance paradigm. Hence, it is possible to recover the $\Lambda$CDM model from the Anton-Schmidt cosmological model. In fact, for $B=0$ and $n=0$, \Cref{eq:Hubble1} reduces to
\begin{equation}
H^2=H_0^2\left(\dfrac{\Omega_{m0}}{a^3}+1-\Omega_{m0}\right)\,,
\label{eq:Hubble LCDM}
\end{equation}
where $1-\Omega_{m0}=\Omega_{\Lambda}$ is the density parameter of the cosmological constant. In fact, in this limit the pressure \Cref{eq:p} becomes a negative constant,
\begin{equation}
P=-\epsilon_c\Omega_{\Lambda}=-\epsilon_{\Lambda}\ ,
\end{equation}
and the equation of state parameter \Cref{eq:w} becomes
\begin{equation}
w=-\dfrac{1-\Omega_{m0}}{\Omega_{m0}a^{-3}+1-\Omega_{m0}}\ ,
\end{equation}
whose present value is
\begin{equation}
w_0=-1+\Omega_{m0}\ .
\end{equation}
As far as the dark energy equation of state  is concerned, for $B=0$ and $n=0$, one has $w_{de}=-1$ (cf. \Cref{eq:dark energy EoS}) which corresponds to the cosmological constant case.


\subsection{Comparison with the CPL model}

A widely used parametrization for the dark energy equation of state is the Chevallier-Linder-Polarski (CPL) model \cite{Chevallier01,Linder03}, which  describes a time-varying dark energy term:
\begin{equation}
w_\text{CPL}(a)=w_0+w_a(1-a)\ .
\label{eq:CPL}
\end{equation}
The above equation represents the first-order Taylor expansion around the present time and allows for deviations from the cosmological constant value $w_0=-1$. Moreover, this model well behaves from high redshifts ($w_\text{CPL}(0)=w_0+w_a$) to the present epoch ($w_\text{CPL}(1)=w_0$). It is possible to relate the constants $w_0$ and $w_a$ to the parameters $n$ and $B$. In fact, expanding \Cref{eq:dark energy EoS} up to the first-order around $a=1$ yields
\begin{equation}
w_{de}\simeq -(1+n+B)-3B^2(1-a)\ ,
\label{eq:Taylor w_de}
\end{equation}
which, once compared with \Cref{eq:CPL}, gives
\begin{equation}
\left\{
\begin{aligned}
&w_0=-(1+n+B)\ , \\
&w_a=-3B^2 .
\end{aligned}
\right .
\end{equation}
When $n=B=0$, we have $w_0=-1$ and $w_a=0$ as in the $\Lambda$CDM model.


\subsection{The onset of cosmic acceleration}

\begin{figure}[h!]
\begin{center}
\includegraphics[width=3.in]{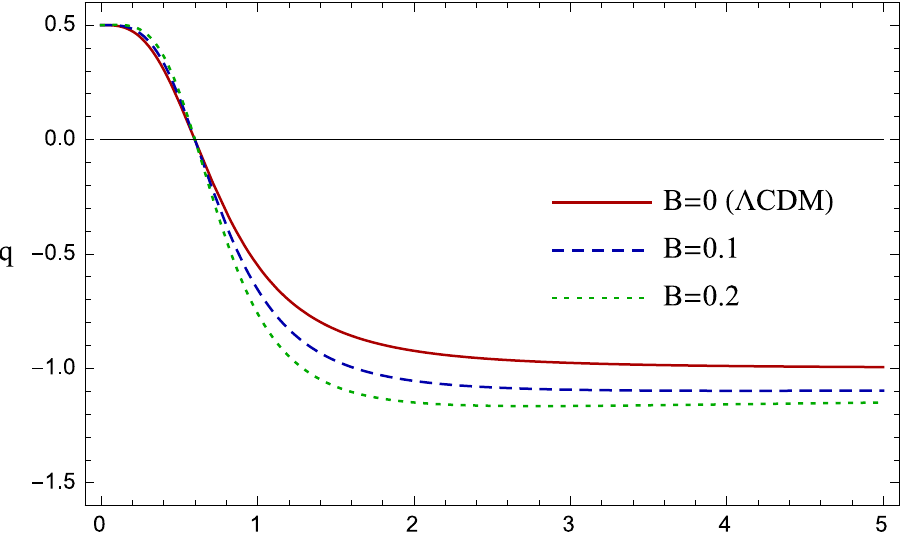}\\
\vspace{0.2cm}
\includegraphics[width=3.in]{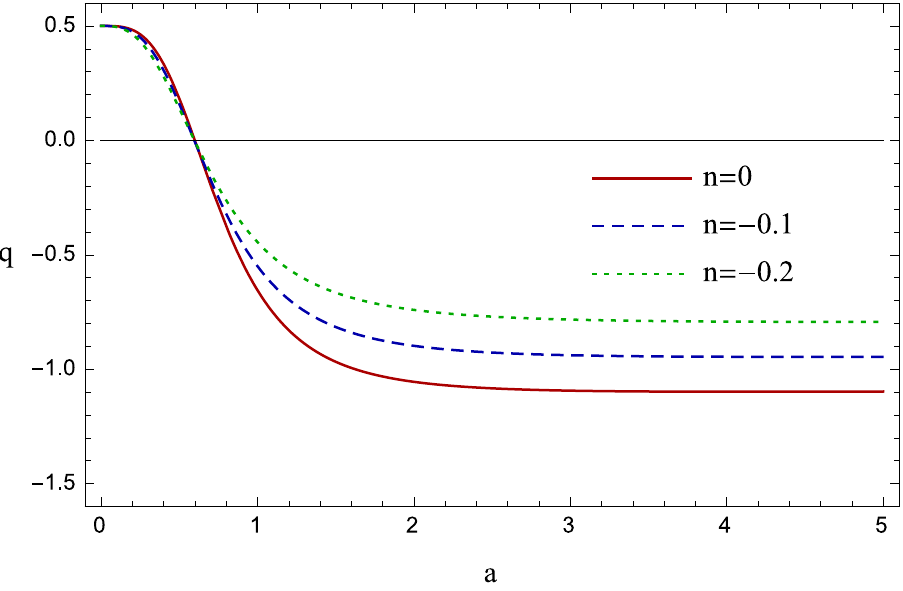}
\caption{Dynamical evolution of the deceleration parameter for different values of $B$ and $n=0$ (\textit{top}), and for different values of $n$ and $B=0.1$ (\textit{bottom}). The matter density is fixed to $\Omega_{m0}=0.3$. The black line separates the region of deceleration $(q>0)$ from the region of acceleration $(q<0)$. $q<-1$ indicates phantom universe.}
\label{fig:q(a)}
\end{center}
\end{figure}

The rate of cosmic expansion is provided by the deceleration parameter \cite{Weinberg72}:
\begin{equation}
q\equiv-\dfrac{\ddot{a} a}{\dot{a}^2}=-\dfrac{\dot H}{H^2}-1\ .
\label{eq:dec param}
\end{equation}
Specifically, the universe is undergoing an accelerated expansion if $-1\leq q< 0$. It is convenient to re-express \Cref{eq:dec param} in terms of the derivative of the expansion rate with respect to the scale factor as
\begin{equation}
q=-\dfrac{a}{H}\dfrac{dH}{da}-1\ .
\end{equation}
Plugging the expression of $H$ given by \Cref{eq:Hubble1} into the above relation, one obtains
\begin{equation}
q=\dfrac{\Omega_{m0}a^{-3}-(1-\Omega_{m0})\left[3(n+B)+3B(3n+2)\ln a+2\right]a^{3n}}{2\left[\Omega_{m0}a^{-3}+(1-\Omega_{m0})(1+3B\ln a)a^{3n}\right]}
\label{eq:q}
\end{equation}
In \Cref{fig:q(a)} we show the analytical behaviour of $q$ as a function of the scale factor for different values of $B$ and $n$, while the matter density parameter is fixed to $\Omega_{m0}=0.3$.
For $n=0$, \Cref{eq:q} reduces to
\begin{equation}
q\Big|_{n=0}=\dfrac{\Omega_{m0}a^{-3}-(1-\Omega_{m0})\left[3B+6B\ln a+2\right]}{2\left[\Omega_{m0}a^{-3}+(1-\Omega_{m0})(1+3B\ln a)\right]}\ ,
\end{equation}
analogous to the LDE models and it converges to $q\rightarrow -1$ when $a\rightarrow \infty$, indicating a de-Sitter phase.
In the limit $n=B=0$, we recover the deceleration parameter of the $\Lambda$CDM model:
\begin{equation}
q_{\Lambda\text{CDM}}=\dfrac{\Omega_{m0}a^{-3}-2(1-\Omega_{m0})}{2\left(\Omega_{m0}a^{-3}+1-\Omega_{m0}\right)}\ .
\end{equation}
The point $a_{acc}$ at which the universe starts accelerating is found for $q=0$, which applied to \Cref{eq:q} gives the condition
\begin{equation}
a^{-3(1+n)}=\dfrac{1-\Omega_{m0}}{\Omega_{m0}}\Big[3(n+B)+3B(3n+2)\ln a+2\Big].
\label{eq:a_acc}
\end{equation}
A straightforward solution is found for $n=-2/3$, for which the above condition reads
\begin{equation}
a_{acc}\Big|_{n=-\frac{2}{3}}=\dfrac{\Omega_{m0}}{3B(1-\Omega_{m0})}\ .
\end{equation}
Further, we can find the condition to get an accelerated universe today. Evaluation of \Cref{eq:q} at the present time yields
\begin{equation}
q_0=\dfrac{\Omega_{m0}-(1-\Omega_{m0})\left[3(n+B)+2\right]}{2} \ ,
\end{equation}
and imposing $-1\leq q_0<0$, one obtains the required condition:
\begin{equation}
\dfrac{\Omega_{m0}-2/3}{1-\Omega_{m0}}< n+B \leq \dfrac{\Omega_{m0}}{1-\Omega_{m0}}\ .
\end{equation}

The transition between the matter and the dark energy eras occurs when the energy densities of both species satisfy $\epsilon_m=\epsilon_{de}$.
Using \Cref{eq:e_m2,eq:e_de1}, the above condition becomes
\begin{equation}
\Omega_{m0}\ a^{-3(n+1)}=(1-\Omega_{m0})(1+3B\ln a)\ .
\label{eq:transition}
\end{equation}
To find the scale factor at the transition $a_{tr}$, we expand the right-hand side of \Cref{eq:transition} in Taylor series up to the first order around $a=1$,
\begin{equation}
(1-\Omega_{m0})(1+3B\ln a)\simeq (1-\Omega_{m0})(1+3B(a-1))\ ,
\end{equation}
inserting it back into \Cref{eq:transition}:
\begin{equation}
\left(\dfrac{\Omega_{m0}}{1-\Omega_{m0}}\right)a^{-3(n+1)}-3Ba\simeq 1+3B\ .
\label{eq:approx transition}
\end{equation}
So, if $n=-2/3$, the solution to \Cref{eq:approx transition} is
\begin{equation}
a_{tr}\Big|_{n=-\frac{2}{3}}\simeq\dfrac{1+3B}{\frac{\Omega_{m0}}{1-\Omega_{m0}}-3B}\ .
\end{equation}
The $\Lambda$CDM limit is thus recovered by setting $n=B=0$ in \Cref{eq:transition}:
\begin{equation}
a_{tr}\Big|_{\Lambda\text{CDM}}={\left(\dfrac{1-\Omega_{m0}}{\Omega_{m0}}\right)}^{-1/3} .
\end{equation}


\subsection{Analyzing the sound speed}

The sound speed plays a key role in theory of perturbations to explain the formation of structures in the universe \cite{perturbations}. It determines the length above which gravitational instability overcomes the
radiation pressure, and the perturbations grow. For an adiabatic fluid, the sound speed is given by
\begin{equation}
c_s^2\equiv \dfrac{\partial P}{\partial \rho}\ .
\label{eq:sound}
\end{equation}
If $c_s$ is comparable to the speed of light, pressure prevents density contrasts to grow significantly, whereas in a matter-dominated universe ($c_s=0$) the gravitational instability on small scales occurs.

\begin{figure}[h!]
\begin{center}
\includegraphics[width=3.2in]{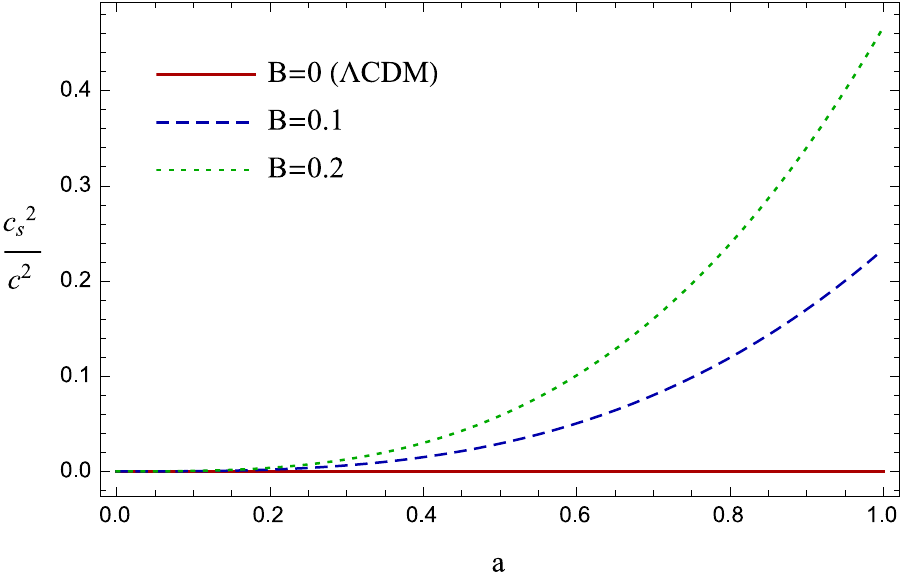}
\caption{Speed of sound as a function of the scale factor for different values of the parameter $B$, while $n=0$.}
\label{fig:cs(a)}
\end{center}
\end{figure}

To calculate the sound speed in terms of the parameters $B$ and $n$, we convert the derivative with respect to the density into the derivative with respect to the scale factor according to
\begin{equation}
\dfrac{\partial \rho}{\partial a}=-\dfrac{3\epsilon_c\Omega_{m0}}{c^2}a^{-4}\ ,
\label{eq:drho_da}
\end{equation}
where we have used \Cref{eq:matter,eq:e_m1,eq:Omega_m}. Therefore, from \Cref{eq:sound} we obtain
\begin{align}
c_s^2&={\left(\dfrac{\partial \rho}{\partial a}\right)}^{-1}\dfrac{\partial P}{\partial a} \nonumber \\
&=c^2\left(\dfrac{1-\Omega_{m0}}{\Omega_{m0}}\right)a^{3(n+1)}\Big[B(1+2n) \nonumber \\
&\hspace{3cm}+n(n+1)(1+3B\ln a)\Big]
\label{eq:c_s}
\end{align}
which turns out to be real only if
\begin{equation}
B(1+2n)+n(n+1)(1+3B\ln a)\geq 0\ .
\end{equation}
In the logotropic limit, \Cref{eq:c_s} becomes
\begin{equation}
\dfrac{c_s^2}{c^2}\bigg|_{n=0}=\dfrac{B}{a^3}\left(\dfrac{1-\Omega_{m0}}{\Omega_{m0}}\right) ,
\end{equation}
which requires $B\geq 0$ for the speed of sound to be real. For $n=B=0$, we have $c_s^2=0$ consistently with the $\Lambda$CDM model. The functional behaviour of the speed of sound for different values of $B$ is displayed on \Cref{fig:cs(a)}.

\section{Observational constraints and experimental limits}
\label{sec:constraints}

In this section, we place observational constraints on our model using the most recent low-redshift cosmological data. To do that, we rewrite the Hubble rate (see \Cref{eq:Hubble1}) in terms of the redshift $z=a^{-1}-1$ as
\begin{equation}
H(z)=H_0\sqrt{\Omega_{m0}(1+z)^3+\dfrac{1-\Omega_{m0}}{(1+z)^{3n}}\left[1-3B\ln (1+z)\right]}\ .
\label{eq:H(z)}
\end{equation}
At this point, we need to interpret the meaning of the reference density $\rho_*$ in \Cref{eq:Anton-Schmidt 2}. The author in \cite{Chavanis15} identified $\rho_*$ with the Planck density $\rho_P$.  The reason of this choice is that the logotropic equation of state applied to dark matter halos shows a constant surface density profile in agreement with observations \cite{Donato09,Saburova14}, providing a very small value of $B$. This implies that the ratio between $\rho_*$ and the dark energy density is huge, of the same order as the ratio between the vacuum energy associated to the Planck density and  $\rho_\Lambda$, which is at the origin of the \textit{cosmological constant problem} \cite{cosm const problem}.
Thus, we set $\rho_*=\rho_P$ and write
\begin{equation}
B=\dfrac{A}{n+1}\left(\dfrac{\rho_{m,0}}{\rho_P}\right)^{-n}\dfrac{1}{\epsilon_{de,0}}\ .
\label{eq:B}
\end{equation}
Since the present dark energy density is
\begin{equation}
\epsilon_{de,0}=-\dfrac{A}{n+1}\left(\dfrac{\rho_{m,0}}{\rho_P}\right)^{-n}\left[\ln\left(\dfrac{\rho_{m,0}}{\rho_P}\right)+\dfrac{1}{n+1}\right] ,
\label{eq:e_de0}
\end{equation}
we soon find that \Cref{eq:B} becomes
\begin{equation}
B=\dfrac{1}{\ln\left(\dfrac{\rho_{P}}{\rho_{m,0}}\right)-\dfrac{1}{n+1}}\ .
\label{eq:B1}
\end{equation}
Recalling the expressions for the present matter and the Planck densities, one has
\begin{equation}
\dfrac{\rho_P}{\rho_{m,0}}=\dfrac{8\pi c^5}{3 \hbar G H_0^2 \Omega_{m0}}\ .
\label{eq:rho_P/rho_m0}
\end{equation}
As a consequence, $B$ is no longer a free parameter of the model, but depends on the fitting parameters $\{H_0,\Omega_{m0}, n\}$ that fully determinate the Hubble rate given by \Cref{eq:H(z)}. We stress that $B$ as expressed in \Cref{eq:B} shows a weak dependence on the value of $n$, since the ratio \eqref{eq:rho_P/rho_m0} is very large. In particular, $B$ takes always small values tending to zero when $n$ approaches $-1$.

In the next section, we present the datasets we used to perform our numerical analysis.


\subsection{Supernovae Ia data}

The largest dataset we consider in our analysis is the JLA sample of SNe Ia standard candles \cite{Betoule14}. This catalogue consists of 740 measurements in the redshift range $0<z<1.3$ and provides model-independent apparent magnitudes at correspondent redshift. The SNe with identical colour, shape and galactic environment are assumed to have on average the same intrinsic luminosity for all redshifts.
Every SN is characterized by a theoretical distance modulus,
\begin{equation}
\mu_{th}(z)=25+5\log_{10}[d_L(z)]
\label{eq:mu}
\end{equation}
where
\begin{equation}
d_L(z) = (1+z)\int_0^z\dfrac{c\ dz'}{H(z')}
\label{eq:dL}
\end{equation}
is the luminosity distance in a flat universe. The distance modulus is modelled as follows:
\begin{equation}
\mu_{obs}=m_B-(M_B-\alpha X_1+\beta C)
\label{eq:dist modulus}
\end{equation}
where $(m_B, X_1, C)$ are the observed peak magnitude in the rest-frame B band, the time stretching of the light-curve and the supernova colour at maximum brightness, respectively. The absolute magnitude $M_B$ is defined based on the host stellar mass $M_{host}$ as
\begin{equation}
M_B=
\begin{cases}
 M, & \text{if}\ M_{host}<10^{10}M_{Sun}\\
  M+\Delta_M, & \text{otherwise}
 \end{cases}
\end{equation}
$M$, $\Delta_M$, $\alpha$ and $\beta$ are nuisance parameters to be determined by a fit to a cosmological model.
We refer the reader to \cite{Betoule14} for the construction of the covariance matrix for the light-curve parameters, which includes both statistical and systematic uncertainties.


\subsection{Observational Hubble Data}

The OHD data represent model-independent direct measurements of $H(z)$. One method is the so-called differential age (DA) method \cite{Jimenez02}, which consist in using passively evolving red galaxies as cosmic chronometers. Once the age difference of galaxies at two close redshifts is measured, one can use the relation
\begin{equation}
\dfrac{dt}{dz}=-\dfrac{1}{(1+z)H(z)}
\end{equation}
to get $H(z)$. In this work, we use a collection of 31 uncorrelated measurements listed in \Cref{tab:OHD}) in the Appendix. In this case, we write the normalized likelihood function as
\begin{equation}
\mathcal{L}_{OHD}=\dfrac{\exp\left[-\dfrac{1}{2}\sum_{i=1}^{31}\left(\dfrac{H_{th}(z_i)-H_{obs}(z_i)}{\sigma_{H,i}}\right)^2\right]} {\left[(2\pi)^{31}\prod_{i=1}^{31} \sigma_{H,i}^2\right]^{1/2}}\ .
\label{eq:likelihood_OHD}
\end{equation}


\subsection{Baryon Acoustic Oscillations}

We use the six model-independent measurements collected and presented in \cite{Lukovic16} (see \Cref{tab:BAO} in the Appendix). These provide the acoustic-scale distance ratio $d_V(z)\equiv r_d/D_V(z)$, where $r_d$ is the comoving sound horizon at the drag epoch, and $D_V(z)$ is a spherically averaged distance measure \cite{Eisenstein05}:
\begin{equation}
D_V(z)= \left[\dfrac{{d_L}^2(z)}{(1+z)^2}\dfrac{c\ z}{H(z)}\right]^{1/3}.
\end{equation}
Being all the measurements uncorrelated, the likelihood function reads
\begin{equation}
\mathcal{L}_{BAO}=\frac{\exp{\left[-\dfrac{1}{2}\sum_{i=1}^6\left(\dfrac{d_V^{th}(z_i)-d_V^{obs}(z_i)}{\sigma_{d_{V,i}}}\right)^2\right]}}{\left[(2\pi)^6\prod_{i=1}^6\sigma_{d_{V,i}}^2\right]^{1/2}}\ .
\label{eq:likelihood_BAO}
\end{equation}


\subsection{Numerical outcomes}

Here, we discuss the results we obtained by combining the datasets discussed above. In this case, we write the joint likelihood function as
\begin{equation}
\mathcal{L}_{joint}=\mathcal{L}_{SN}\times \mathcal L_{OHD}\times \mathcal L_{BAO}\ .
\end{equation}
To test our model, we performed the Markov Chain Monte Carlo (MCMC) algorithm for parameter estimation using the \textit{emcee} software package \cite{Foreman-Mackey13}, which is an implementation of the affine-invariant ensemble sampler of \cite{Goodman10}.
We assumed uniform priors for the parameters, listed in \Cref{tab:priors}.
\begin{table}[h!]
\begin{center}
\renewcommand{\arraystretch}{1.3}
\begin{tabular}{c c  }
\hline
\hline
Parameter & Prior \\
\hline
$H_0$  & $(50,100)$ \\
$\Omega_{m0}$ & $(0,1)$ \\
$n$ & $(-1,1)$\\
$M$ & $(-19.5,-18.5)$ \\
$\Delta_M$  & $(-0.1,0)$ \\
$\alpha$  & $(0,0.3)$ \\
$\beta$  & $(0,5)$ \\
$r_d$  & $(140,160)$ \\
\hline
\hline
\end{tabular}
\caption{Priors for parameters estimate in the MCMC numerical analysis.}
 \label{tab:priors}
\end{center}
\end{table}

\begin{figure*}[h!]
\begin{center}
\includegraphics[width=1\textwidth]{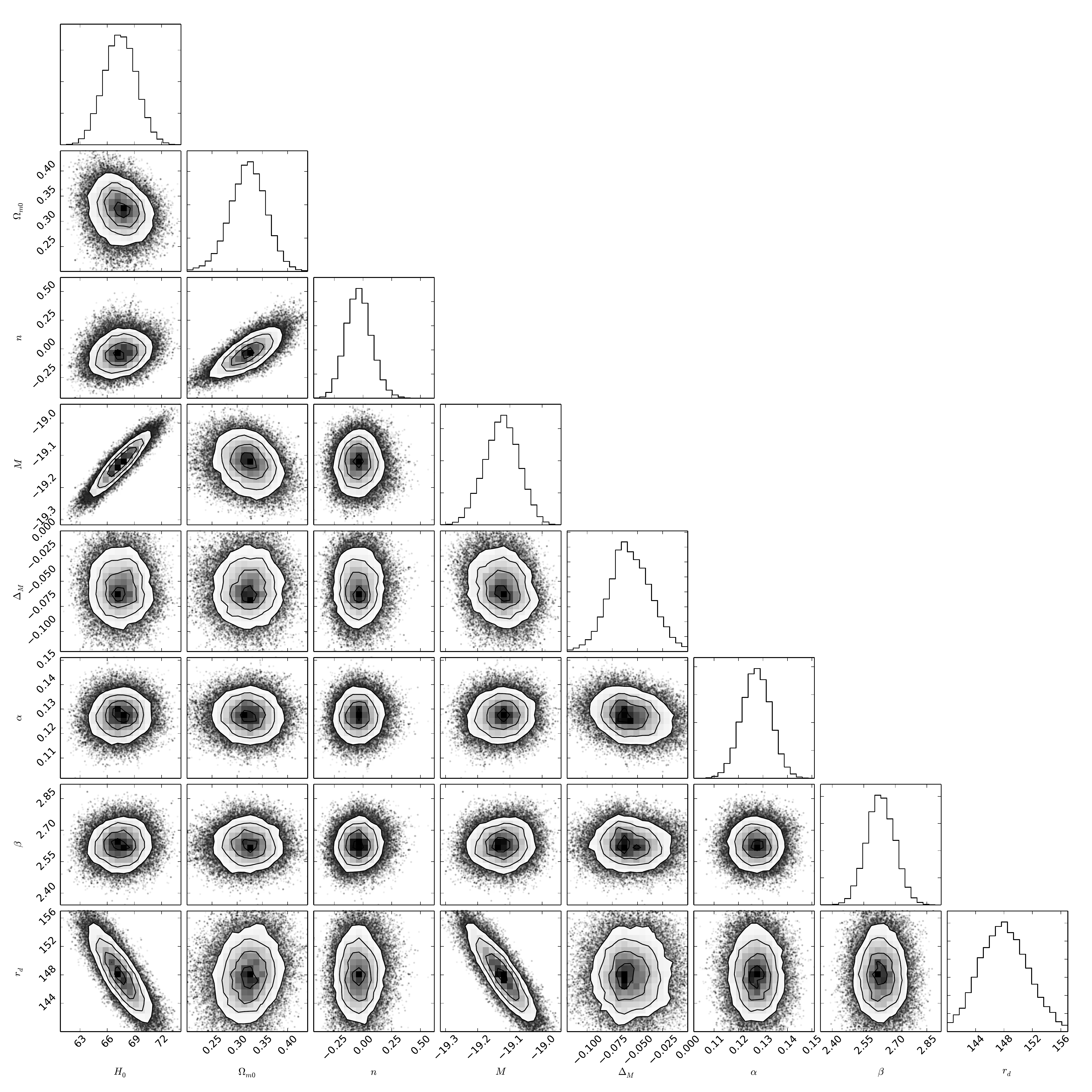}
\caption{Marginalized contours at 68\%, 95\%, 99\% confidence levels and posterior distributions for the parameters of the Anton-Schmidt cosmological model.}
\label{fig:contours}
\end{center}
\end{figure*}

Our results are shown in \Cref{tab:results}, and \Cref{fig:contours} shows the probability contours for the parameters of the model.
The high relative error on the estimate of $n$ from the joint fit entails that our model is undistinguishable from the $\Lambda$CDM model $(n=0)$ at the $2\sigma$ confidence level.
Our 2.7\% estimate of the Hubble constant results to be consistent within $1\sigma$ with the CMB estimate of the Planck collaboration \cite{Planck15}, $H_0=67.31 \pm 0.96$.  Also the present matter density parameter is consistent with $\Omega_{m0}=0.315 \pm 0.013$ found by Planck, although the relative error exceeds 10\%.

\begin{widetext}
\hspace{1cm}\begin{table}[h!]
\begin{center}
\setlength{\tabcolsep}{1.2em}
\renewcommand{\arraystretch}{1.5}
\begin{tabular}{c c c c c c }
\hline
\hline
Parameter & $H_0$+SN & OHD & BAO & SN+OHD+BAO \\
\hline
$H_0$ & 70 & $64.53^{\ +8.86}_{\ -6.81}$  & $62.37^{\ +4.09}_{\ -3.80}$  & $65.67^{\ +1.75}_{\ -1.78}$  \\
$\Omega_{m0}$ & $0.107^{\ +0.111}_{\ -0.128}$ & $0.242^{\ +0.065}_{\ -0.061}$  & $0.272^{\ +0.051}_{\ -0.056}$  & $0.286^{\ +0.034}_{\ -0.036}$ \\
$n$ & $-0.382^{\ +0.239}_{\ -0.170}$ & $-0.251^{\ +0.699}_{\ -0.590}$  & $-0.336^{\ +0.315}_{\ -0.283} $ & $-0.147^{\ +0.113}_{\ -0.107}$  \\
$M$ & $-19.07^{\ +0.03}_{\ -0.02}$ & - & - & $-19.18^{\ +0.05}_{\ -0.06}$  \\
$\Delta_M$  & $-0.075^{\ +0.021}_{\ -0.021}$ & - & - & $-0.077^{\ +0.021}_{\ -0.019}$ \\
$\alpha$  & $0.121^{\ +0.006}_{\ -0.006}$ & - & - & $0.121^{\ +0.006}_{\ -0.006}$  \\
$\beta$  & $2.559^{\ +0.067}_{\ -0.068}$ & - & - & $2.565^{\ +0.068}_{\ -0.066}$  \\
$r_d$  & - & - & $142.9^{\ +6.9}_{\ -6.6}$ & $144.6^{\ +3.5}_{\ -3.3}$ \\
\hline
\hline
\end{tabular}
\caption{68\% confidence level parameter constraints resulting from the MCMC analysis of different data. The SN data alone are insensible to the value of $H_0$, which has been fixed to 70. $H_0$ values are expressed in units of km/s/Mpc, and $r_d$ values in units of Mpc.}
 \label{tab:results}
\end{center}
\end{table}
\end{widetext}

\begin{figure}[h!]
\begin{center}
\includegraphics[width=3.2in]{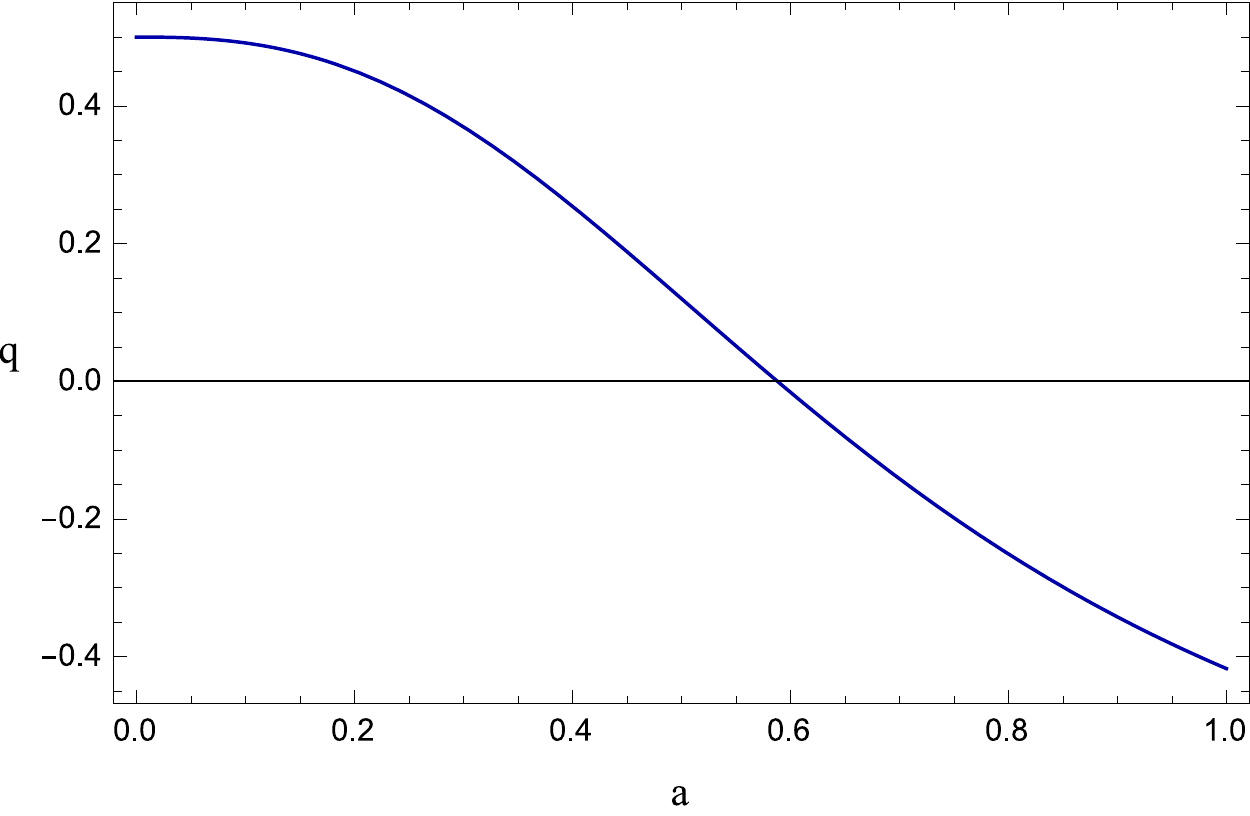}
\caption{Deceleration parameter as function of the scale factor resulting from the best-fit values of our joint analysis. The intersection of the curve with the black line corresponds to the point at which the universe starts accelerating.}
\label{fig:q}
\end{center}
\end{figure}

Using the best-fit values of the parameters $\{H_0,\Omega_{m0}, n\}$ from the joint analysis, it is possible to calculate the parameter $B$ by means of \Cref{eq:B1,eq:rho_P/rho_m0}. In particular, we obtain
\begin{equation}
B=3.54 \times 10^{-3}\ ,
\label{eq:B value}
\end{equation}
in agreement with the value predicted in \cite{Chavanis17}. For such a value of $B$, one can calculate from \Cref{eq:a_acc} the point when the universe starts accelerating:
\begin{equation}
a_{acc}=0.59\ ,
\end{equation}
or, equivalently in terms of redshift,
\begin{equation}
z_{acc}=0.70\ .
\end{equation}
This result is in agreement with the recent estimate $z_{acc}=0.67^{+0.10}_{-0.08}$ found in \cite{transition}. The functional behaviour of the deceleration parameter $q(a)$ is displayed on \Cref{fig:q}.

\clearpage


\section{Final outlooks and perspectives}
\label{sec:conclusions}

In this paper, we proposed a new class of dark energy models, in which dark energy emerges as a consequence of the kind of barotropic factor involved for matter. We considered in Solid State Physics the case supported by isotropic deformations into crystalline solids, in which the pressure can naturally take negative values. This fact avoids to postulate a priori a negative dark energy equation of state and moreover adapts well to matter. So that, considering matter only, we model its equation of state through the Anton-Schmidt's equation of state.
Under the hypothesis that $V\propto \rho^{-1}$, we soon found that the corresponding pressure is proportional to the logarithm of universe's volume, i.e. to the density itself. We demonstrated that a universe made of one single fluid with such a property can explain acceleration without the need of a cosmological constant. Our framework depends upon two free parameters, namely $n$ and $B$, intimately correlated with  the Gr\"uneisen parameter and the bulk modulus characterizing the typology of fluid here involved.
We noticed that the acceleration process extends a class of dark energy paradigms dubbed \emph{logotropic} dark fluids recently developed in the literature.
We discussed the connections between these models and our model, and we showed the differences between our treatment and the concordance paradigm. Further, we related the free parameters of the CPL approach to our treatment, finding out the analogies between the two landscapes.

We highlighted the most relevant properties of the new equation of state applied to background cosmology. To fix bounds on the free parameters of our model, we employed supernova data surveys, BAO compilations and differential age measurements. We thus performed Monte Carlo computing technique implemented with the affine-invariant ensemble algorithm.
In so doing, we were able to show that our alternative model candidates as a consistent alternative to dark energy and to the $\Lambda$CDM model. In particular, the concordance paradigm falls inside our approach as a limiting case.

In future works, we will analyze some open challenges of our approach. First, we will focus on better understanding the physics inside $\rho_*$, comparing our model with dark matter distributions. Second, we will check the goodness of our working scheme when early phases are taken into account. Last but not least, we will clarify whether the Anton-Schmidt's equation of state is also able to produce inflationary stages of the universe, without the need of scalar fields in inflationary puzzle. Finally, we will check whether the model is capable of overcoming and passing additional experimental tests at different redshift regimes.

\begin{acknowledgements}
This paper is based upon work from COST action CA15117  (CANTATA), supported by COST (European Cooperation in Science and Technology). S.C. acknowledges the support of INFN (iniziativa specifica QGSKY).
R.D. is grateful to Dr. Giancarlo de Gasperis for useful suggestions on the Monte Carlo analysis. O.L. warmly thanks the National Research Foundation (NRF) of South Africa for financial support.
\end{acknowledgements}

\appendix*
 \section{Experimental data compilations}
 \label{sec:appendix}
\vspace{-0.5cm}
In this appendix, we list the Hubble measurements and baryon acoustic oscillations data used in this work to get bounds over the free parameters of our model.
\begin{table}[h]
\footnotesize
\begin{center}
\begin{tabular}{c c c }
\hline
\hline
 $z$ &$H \pm \sigma_H$ &  Ref. \\
\hline
0.0708	& $69.00 \pm 19.68$ & \cite{Zhang14} \\
0.09	& $69.0 \pm 12.0$ & \cite{Jimenez02} \\
0.12	& $68.6 \pm 26.2$ & \cite{Zhang14} \\
0.17	& $83.0 \pm 8.0$ & \cite{Simon05} \\
0.179 & $75.0 \pm	4.0$ & \cite{Moresco12} \\
0.199 & $75.0	\pm 5.0$ & \cite{Moresco12} \\
0.20 &$72.9 \pm 29.6$ & \cite{Zhang14} \\
0.27	& $77.0 \pm 14.0$ & \cite{Simon05} \\
0.28	& $88.8 \pm 36.6$ & \cite{Zhang14} \\
0.35	& $82.1 \pm 4.85$ & \cite{Chuang12}\\
0.352 & $83.0	\pm 14.0$ & \cite{Moresco16} \\
0.3802	& $83.0 \pm 13.5$ & \cite{Moresco16}\\
0.4 & $95.0	\pm 17.0$ & \cite{Simon05} \\
0.4004	& $77.0 \pm 10.2$ & \cite{Moresco16} \\
0.4247	& $87.1 \pm 11.2$  & \cite{Moresco16} \\
0.4497 &	$92.8 \pm 12.9$ & \cite{Moresco16}\\
0.4783	 & $80.9 \pm 9.0$ & \cite{Moresco16} \\
0.48	& $97.0 \pm 62.0$ & \cite{Stern10} \\
0.593 & $104.0 \pm 13.0$ & \cite{Moresco12} \\
0.68	& $92.0 \pm 8.0$ & \cite{Moresco12} \\
0.781 & $105.0 \pm 12.0$ & \cite{Moresco12} \\
0.875 & $125.0 \pm 17.0 $ & \cite{Moresco12} \\
0.88	& $90.0 \pm 40.0$ & \cite{Stern10} \\
0.9 & $117.0 \pm 23.0$ & \cite{Simon05} \\
1.037 & $154.0 \pm 20.0$ & \cite{Moresco12} \\
1.3 & $168.0 \pm 17.0$ & \cite{Simon05} \\
1.363 & $160.0 \pm 33.6$ & \cite{Moresco15} \\
1.43	& $177.0 \pm18.0$ & \cite{Simon05} \\
1.53	& $140.0	\pm 14.0$ & \cite{Simon05} \\
1.75	 & $202.0 \pm 40.0$ & \cite{Simon05} \\
1.965& $186.5 \pm 50.4$ & \cite{Moresco15} \\
\hline
\hline
\end{tabular}
\caption{Differential age $H(z)$ data used in this work. The Hubble rate is given in units of km/s/Mpc.}
 \label{tab:OHD}
\end{center}
\end{table}
\vspace{-1.2cm}
\begin{table}[h]
\begin{center}
\footnotesize
\begin{tabular}{ c c c c }
\hline
\hline
 $z$ &$d_V \pm \sigma_{d_V}$ &  Survey &  Ref. \\
\hline
0.106 & 0.336 $\pm$  0.015 &6dFGS   &  \cite{Beutler11}\\
0.15 & 0.2239 $\pm$ 0.0084 & SDSS DR7 & \cite{Ross15} \\
0.32 &  0.1181 $\pm$ 0.0023 & BOSS DR11 & \cite{Anderson14} \\
0.57 & 0.0726 $\pm$ 0.0007 & BOSS DR11 & \cite{Anderson14} \\
2.34 & 0.0320 $\pm$ 0.0016 & BOSS DR11 & \cite{Delubac15} \\
2.36 &  0.0329 $\pm$ 0.0012 & BOSS DR11 & \cite{Font-Ribera14}\\
\hline
\hline
\end{tabular}
\caption{BAO data used in this work.}
 \label{tab:BAO}
\end{center}
\end{table}

\end{document}